# Temperature of systems out of thermodynamic equilibrium

J.-L. Garden[a)], J. Richard and H. Guillou

*Institut Néel, CNRS et Université Joseph Fourier, BP 166, 38042 Grenoble Cedex 9, France.*

**Abstract**

Two phenomenological approaches are currently used in the study of the vitreous state. One is based on the concept of fictive temperature introduced by Tool [Jour. Research Nat. Bur. Standards **34**, 199 (1945)] and recently revisited by Nieuwenhuizen [Phys. Rev. Lett. **80**, 5580 (1998)]. The other is based on the thermodynamics of irreversible processes initiated by De Donder at the beginning of the last century [*L'Affinité* (Gauthier-Villars, Paris, 1927)] and recently used by Möller and co-workers for a thorough study of the glass transition [J. Chem. Phys. **125**, 094505 (2006)]. This latter approach leads to the possibility of describing the glass transition by means of the freezing-in of one or more order parameters connected to the internal structural degrees of freedom involved in the vitrification process. In this paper, the equivalence of the two preceding approaches is demonstrated, not only for glasses, but in a very general way for any system undergoing an irreversible transformation. This equivalence allows the definition of an effective temperature for all systems departed from equilibrium generating a positive amount of entropy. In fact, the initial fictive temperature concept of Tool leads to the generalization of the notion of temperature for systems out of thermodynamic equilibrium, for which glasses are just particular cases.

[a)]*Electronic mail:* jean-luc.garden@grenoble.cnrs.fr

## I. INTRODUCTION

The mysterious glass transition is mainly due to the gradual freezing of one or several degrees of freedom when the temperature of a system is decreased below its temperature of crystallization (for example a glass-forming liquid or a polymer). However, the understanding of the thermodynamic basis of the glass transition and the relationship existing between thermodynamics and dynamics (or kinetics) seem to be still under debate.[1] While the glass transition occurs, the system is for a time out of its state of equilibrium until it reaches a new equilibrium state when it is completely frozen. To our knowledge, the first attempt to envisage the glass transition in considering the progressive freezing-in of the internal modes of a system under a thermodynamic point of view has been done by Simon and latter by Bernal, Jones, Davies, Prigogine and Defay.[2-5] At the same time, in order to conciliate theory and experiments on the glass transition, Tool has invented a useful phenomenological concept called the fictive temperature.[6,7] This parameter, which has the same physical dimension as a temperature can describe how far a system is departed from its initial equilibrium state. The fictive temperature is nowadays extensively used theoretically and experimentally in studies on the vitreous state or on systems with slow internal dynamics.[8,9] Nieuwenhuizen has recently developed a new thermodynamics based on the fictive temperature.[10] Using this effective temperature as additional thermodynamic variable, he has explained most of the "old questions" concerning the glass transition.[11,12] Nevertheless, what does this temperature really means and how it is experimentally accessible is still under discussion. Nowadays, the notion of temperature for systems departed from thermodynamic equilibrium is still not exactly clarified, especially from a statistical physics point of view, and the ongoing literature on the subject is very abundant.[13-15] A recent review on the subject has been written by Casas-Vázquez and Jou.[16] In the low temperatures field, however, it is demonstrated that different kinds of particles such as phonons and electrons can be transiently thermally decoupled within a single material and thus, can be brought at different temperatures as measured for instance by means of electronic noise.[17] Recently, it has been finely demonstrated that at such low temperatures the transfer of heat is indifferently carried out by phonons, electrons or, when the electron-phonon interaction is frozen, by photons.[18] However, what does the classical statistical temperature exactly mean when the system undergoes a glass transition? Does the current statistical physics even make sense for vitreous systems for which the fundamental ergodic hypothesis seems to be not valid anymore?[19] Likewise, does the notion of temperature has a real signification for macroscopic systems out of thermodynamic equilibrium? These questions are assessed in this paper when the fictive temperature is envisaged by means of the classical theory of irreversible processes.

The paper is decomposed into six sections. After the introduction, in the section II we present the current literature around which the present paper is based. Especially, in a first part the macroscopic non-equilibrium thermodynamics based on De Donder's precursor works is briefly presented, and in a second part the thermodynamics recently developed by Nieuwenhuizen is provided. A brief survey of the initial fictive temperature concept of Tool under which the thermodynamics of Nieuwenhuizen is based is presented. In the section III, we argue that the two approaches are equivalent and particularly that the effective temperature and the order parameter are two variables which are not independent of each others. We give several relationships connecting these two variables. For that purpose, a simple thermal model based on the Fourier's equation on the propagation of heat is employed and a comparison of macroscopic quantities such as the rate of production of entropy and the heat capacity are used in order to establish the connection between the two approaches. In the section IV, we used this connection in order to show that a supplementary thermodynamic macroscopic temperature can be defined for all systems undergoing an irreversible process. The particular case of glasses is discussed. In the section V, we envisage experiments which can reveal the existence of an effective temperature as defined in the previous section. The last section VI is a concluding summary.

## II. TWO PHENOMENOLOGICAL APPROACHES OF THE GLASS TRANSITION

**A. Macroscopic non-equilibrium thermodynamics based on De Donder precursor works**

Among the thermodynamic coefficients, the heat capacity is likely one of the best adapted in describing the glass transition. This thermodynamic quantity, indeed, makes evident how internal degrees of freedom are thermally activated within a body. Another usual property in considering systems departed from thermodynamic equilibrium, for example during aging of a glass, is the so-called entropy production. This property is always linked to an amount of heat dissipated inside the system when sub-microscopic friction takes place in the volume delimited by the body due to the slow internal reorganisation of some degrees of freedom. Prior to describe succinctly the phenomenology of the glass transition, let us recall briefly the expression of the two principles of thermodynamics following the classical non-equilibrium thermodynamics. Owing to the Belgian school of thermodynamics, the first law is written in making apparent a supplementary variable, the so-called order parameter $\xi$, which characterizes a specific internal degree of freedom involved in the change of the order of the system during the transformation:[20,21]

$$dU = TdS + \delta W - A d\xi \tag{1}$$

where $A=-\partial G/\partial \xi)_{T,P}$ is the affinity of thermodynamic transformation, $G$ being the Gibbs free energy. Although initially the above equation has been derived by De Donder only in the case of chemical reactions, where $\xi$ represents the degree of advance of the reaction (and $A$ the affinity of the reaction), this variable has been next generalized by Prigogine, De Groot and Mazur in order to characterize any irreversible processes.[22,23] Today, any irreversible process can be described by fluxes of ordering parameters with associated affinities. Recently, Rubi, Pérez-Madrid, Vilar and Reguera have developed a very clear and interesting general mesoscopic non-equilibrium thermodynamics based on the generalization of internal degrees of freedom to any irreversible processes extended outside the local equilibrium assumption.[24-26] Following the classical thermodynamics of irreversible processes, the internal creation of entropy is always expressed by the product of the thermodynamic force ($A/T$ for instance) by the advance of the conjugated thermodynamic order parameter ($d\xi$ for instance):

$$d_i S = \frac{A}{T} d\xi \qquad (2)$$

Taking into account this notion of order parameter, the heat capacity of a system at constant pressure measured for example during a calorimetric experiment is written:[21]

$$C_p = \frac{\delta Q}{dT} = C_\xi + \partial H/\partial \xi)_T \frac{d\xi}{dT} \qquad (3)$$

$\delta Q$ is the amount of heat (positive or negative) exchanged between the system and the thermal surroundings (thermal bath); $dT$ is the temperature variation effectively recorded by the experimentalist; $H$ is the enthalpy of the system; $C_\xi = \partial H/\partial T)_\xi$ is the heat capacity at constant order parameter. $\partial H/\partial \xi)_T$ is called the isothermal heat of the transformation due to the advancement of the order parameter. The affinity of the process (driving force) being a state function, it is differentiable with respect to the temperature and the order parameter as follows:

$$dA = \partial A/\partial T)_\xi dT + \partial A/\partial \xi)_T d\xi \tag{4}$$

It can be demonstrated that the first partial derivative coefficients of the affinity is $\partial A/\partial \xi)_T = -\partial^2 G/\partial \xi^2)_T$ and the second $\partial A/\partial T)_\xi = \partial S/\partial \xi)_T$. So, knowing that when the system is out of thermodynamic equilibrium the time becomes a preponderant variable, then the heat capacity can be explicitly written:

$$C_P = \frac{\delta Q/dt}{dT/dt} = C_\xi + \partial H/\partial \xi)_T \frac{d\xi/dt}{dT/dt} \tag{5}$$

Replacing the rate of the order parameter in the above equation by means of (4), the general expression of the heat capacity of a system departed from equilibrium was first obtained by Prigogine and Defay:[27]

$$C_P = C_\xi + \frac{[\partial H/\partial \xi)_T]^2}{T\partial^2 G/\partial \xi^2)_T} + \frac{A\partial H/\partial \xi)_T}{T\partial^2 G/\partial \xi^2)_T} - \frac{\partial H/\partial \xi)_T}{\partial^2 G/\partial \xi^2)_T}\frac{dA/dt}{dT/dt} \tag{6}$$

This expression emphasizes the presence of the affinity in the heat capacity. More important is the ratio of the affinity rate on the temperature rate which defines the internal response of the system departed from equilibrium to an experimentally imposed temperature variation of well-fixed rate.

From the previous equations, it is possible to briefly describe the phenomenology of the glass transition. Actually, we observe from (4) that it is possible to be in a situation where for a given temperature variation $\Delta T$, over a time scale $\Delta t$, the order parameter has had not enough time to evolve over this time scale ($\Delta \xi = 0$). In this case, the affinity variation is only due to the temperature variation, or in other words, the temperature rate is infinitely high as compared to the order parameter rate. Injecting this value of the affinity rate in the general expression (6), the equality $C_P = C_\xi$ is obtained. It is the heat capacity of a system for which the parameter $\xi$ has been frozen (glassy state). The system is in the so-called arrested equilibrium over the time scale of observation like in the case of a glass transition. The configurational heat capacity, which is the contribution to the total heat capacity of the structural degree of freedom, given by the three last terms of (6), does not exist anymore. In the other extreme case, if the temperature rate is so slow that the order parameter takes its equilibrium value at each instant and each temperature, the configurational heat capacity is maximum. That is to say that not only all the fast internal degrees of freedom are energetically excited, but also that the internal degree of freedom concerned by the order parameter is also excited (the order of the system has changed). In this case, it is possible to show that the heat capacity can be expressed as follow:[27]

$$C_{eq} = C_\xi + \frac{[\partial H / \partial \xi)_T^{eq}]^2}{T \partial^2 G / \partial \xi^2)_T^{eq}} \tag{7}$$

where the second term of the right-hand side outlines the "energetic contribution" of the considered order parameter to the heat capacity (often called the enthalpy peak). In fact, in a real temperature cooling experiment on a glass-forming liquid the situation just above is impossible to obtain, even for slow temperature scans, because the relaxation time of the order parameter becomes so high in the annealing temperature range that progressively the order parameter rate tends toward zero bringing the system in the first case discussed above (glassy state). The most observable effect is a decrease of the glass transition temperature $T_g$. This latter temperature can be defined from the inflection point in the $C_P(T)$ jump during the transition for example. At this stage, we would like to discuss one simplification that we will consider in the whole paper. For example, we will always consider one single order parameter in describing basically the α-relaxation process which is the slowest observable structural relaxation in the annealing region. The possibility of using this assumption is, however, still under debate (see for example references [10,28,29]) and it is not our aim to enter in the details of the glass transition feature, but just to give phenomenological models which could improve our thermodynamic understanding of the glass transition phenomenon and especially to discuss the notion of temperature for glasses. For very complete and interesting reviews on glass transition, see the following references [30] to [34].

**B. Phenomenological approaches of the fictive and effective temperatures**

Tool has recognized himself that the time variable is of high importance in glass transition studies.[6,7] In the glass transformation range, when the time necessary for internal configurational readjustments exceeds the time taken by the temperature to undergo a sudden change, the system is for an instant in a thermodynamic non-equilibrium state with regard to some of its internal degrees of freedom. As Tool did, it is thus possible to define one extra parameter, called the fictive temperature, which has a dimension of a temperature, and which takes into account the slow evolution of some internal degrees of freedom inside the system. He defined the fictive temperature, near equilibrium, by means of a simple linear first-order differential equation:[7]

$$\frac{dT_f}{dt} = K_T(T - T_f) \tag{8}$$

where $T_f$ is the fictive temperature. $T$ is the classical temperature. $K_T$ is a positive coefficient generally exponentially depending on $T$. Tool has initially given a simple expression for this coefficient: $K_T = K \exp(T/k)$ with $k$ the Twyman's constant and K the value of $K_T$ for $T = 0$ (see ref. [7]). From (8), it can be seen that if a temperature $T_0$ is maintained constant, the fictive temperature $T_f$ relaxes toward $T_0$ following the relaxation time constant $\tau = 1/K_T$. At a first order, this decay effect is connected to aging, annealing or structural recovery while it is experimentally observed during the measurement of physical properties such as the volume or the enthalpy. If the temperature is cooled following a temperature ramp crossing the glass transition, $\tau$ tends rapidly towards a very high value, the fictive temperature rate thus becomes also progressively close to zero and the system becomes finally in an arrested-equilibrium state (freezing-in of the slow modes or divergence of the viscosity). In this latter case, the system is a glass for which a frozen fictive temperature can be estimated from calorimetric measurement for example[35]. It defines actually the equilibrium temperature of the system when it just became to be frozen at a previous instant. We will show in the following that this frozen fictive temperature, which is the conventional parameter used in glass transition, is not a thermodynamic temperature. In contrast, the initial definition of the fictive temperature according to Tool (equation (8)) is a thermodynamic temperature in the classical meaning.

In a recent version due to Nieuwenhuizen, based nevertheless on Tool's approach, the fictive temperature is regarded as the real temperature of the slow relaxing modes within a system in the annealing range.[10-12,36-38] Nieuwenhuizen defines the fictive temperature (mostly called effective temperature in his works) as an extra thermodynamic parameter included directly in the heat exchange appearing in the first law of thermodynamics $dU = \delta Q + \delta W$, where the heat exchange is separated in two contributions:

$$\delta Q = TdS_{ep} + T_e dI \qquad (9)$$

$T$ and $T_e$ are the classical and effective temperatures respectively. $dS_{ep}$ and $dI$ are the two components of the total entropy variation of the system:

$$dS = dS_{ep} + dI \qquad (10)$$

Nieuwenhuizen has called $S_{ep}$ the equilibrium part of the system entropy and $I$ the configurational part or complexity.[10] In the case of glasses, $I$ (sometimes called residual entropy) is connected to the number of macroscopic glassy states that a system can potentially have, each of these states being dependant on the way (e.g cooling temperature ramp) for obtaining them. For example, for one temperature cooling rate, this part of the entropy disappeared in the glass transition region because the system is frozen in one defined macroscopic state.[12] From these two fundamental equations, Nieuwenhuizen solves several paradoxes concerning the glass transition phenomenon. For example, this approach has allowed a better understanding of the non unity of the Prigogine-Defay ratio and has led to the modification of the Ehrenfest relations and to the modification of the Maxwell relations when other works than the mechanical one are taken into account (see for details references [10] and [12]). Particularly, with the separation of the heat into an equilibrium and a non-equilibrium part, he has derived the following expression for the heat capacity:[11]

$$C_P = \frac{\delta Q}{dT} = T\frac{dS_{ep}}{dT} + T_e\frac{dI}{dT} \qquad (11)$$

## III. EQUIVALENCE OF THE NON-EQUILIBRIUM THERMODYNAMICS OF DE DONDER/PRIGOGINE/DEFAY AND THE EFFECTIVE TEMPERATURE FRAMEWORK OF TOOL/NIEUWENHUIZEN

### A. Fourier's model and the effective temperature

To our knowledge, no connections have been already established between the $\xi$-order parameter and the fictive temperature concept. Let us, however, mention the work of Baur on thermodynamic relaxation in the glass transition region by means of thermodynamics of irreversible processes, although the author used $\xi$ and $T_f$ as two independent variables in describing the glass transition.[39,40] Let us also mention the very interesting and recent work of Möller, Schmelzer and Gutzow who have used the non-equilibrium thermodynamics of De Donder/Prigogine/Defay for the study of the glass transition.[29,41] In one of their paper concerning the Prigogine-Defay ratio, the authors conclude that a comparison must be of interest between the classical non-equilibrium thermodynamics and that of Nieuwenhuizen without pushing further the reasoning.[29] The aim of this present section is to demonstrate that $\xi$ and the effective temperature of Nieuwenhuizen $T_e$, although being known as two parameters employed in different frameworks, are actually not independent of each other. But, in order to establish a clear connection between the two approaches, we have firstly to push further the thermodynamics of Nieuwenhuizen by using a very simple thermal model depicted in the figure 1. Indeed, since Nieuwenhuizen states that $T_e$ is the temperature of the slow relaxing modes during the glass transition, let us describe, by analogy with the Fourier's law, the irreversible exchange of heat between two parts having different temperatures. In the sketch of the figure 1, a thermodynamic homogenous discrete system is composed of two different sub-parts; one is linked to the thermal bath and has an equilibrium classical thermodynamic temperature $T$; the other part is thermally connected to this first part by means of a thermal conductance $K_i$, and has an effective temperature $T_e$. This latter system represents only the configurational degrees of freedom. This picture depicts the thermal coupling between the fast modes of classical temperature $T$ and the slow modes of temperature $T_e$. This minimalist model enables us to describe a time-scale separation between different internal degrees of freedom constituting a whole system submitted to a time dependent external perturbation. We have, however, to bear in mind that in the figure 1, instead of considering an heat exchange between two parts of different temperatures in a classical three dimensional space as it is usually the case for the Fourier's law, we have to imagine that this heat exchange takes place between different degrees of freedom of different temperatures following a virtual axis represented by the advance of the process occurring at any point in the volume of the entire system. The advancement of $\xi$ in the course of time defines this virtual axis. In this case, the order parameter $\xi$ is the "internal macroscopic coordinate" of the whole system such as used in the energy landscape theory.[42] To consider any irreversible transformation as a diffusion process inside a virtual internal space of coordinate has already been envisaged (see the following references [24] and [43] for diffusion of internal degrees of freedom by means of non-equilibrium thermodynamics). In our case, the analogy is pushed farther because we consider a diffusion of heat between the effective temperature and the classical temperature occurring at any points inside the volume delimiting the entire system. Under this circumstance, the classical version of the Fourier's law in stationary conditions, $dQ/dt = K\Delta T$, is a transport equation which can be obtained directly from the thermodynamics of irreversible processes like all transport equations (Fick's law, Ohm's law, etc…). The getting of transport equations from the physics of irreversible processes requires an underlying assumption, due to Onsager, defining linear relations between thermodynamic fluxes and forces.[22,23] In the case of our minimalist model, where for simplicity we have used discrete sub-systems, the thermodynamic force involved in the system (generalized affinity) is:

$$\Delta\left(\frac{1}{T}\right) = \frac{1}{T_e} - \frac{1}{T} \tag{12}$$

Owing to Onsager, the thermodynamic flux of the conjugated variable (heat flux) is simply proportional to the force:

$$\frac{dQ}{dt} = L\left(\frac{1}{T_e} - \frac{1}{T}\right) = L\left(\frac{T - T_e}{T_e T}\right) \tag{13}$$

where $L$ is the kinetic coefficient of Onsager. This is the Fourier's law where $K_i = L/T_e T$. In a stationary state, the preceding picture gives rise to a permanent production of entropy written as the product of the flux with the force:

$$\sigma_i = \frac{d_i S}{dt} = \frac{K_i (T - T_e)^2}{T_e T} \tag{14}$$

For non-stationary situations, there is an irreversible process due to the relaxation of the effective temperature toward the equilibrium temperature of the system and where the heat capacity of the slow mode, $C_i$, plays a role. In this case, this relaxation process is driven by the classical dynamic equation of heat where the relaxation time constant $\tau = C_i / K_i$ governs the dynamic of the heat exchanges.

**B. Fourier's model and the effective temperature in Nieuwenhuizen's approach**

The second law of thermodynamics states that the total entropy variation during an irreversible transformation is greater than the entropy variation due only to reversible exchanges of heat between a system and its surrounding. This second principle can be written in the form of equality when making apparent the positive entropy produced by non-equilibrium processes occurring within the system (entropy source term):

$$dS = d_e S + d_i S \tag{15}$$

where $d_e S$ is the external contribution due only to the quantity of heat exchange between the system and the surrounding $d_e S = \delta Q / T$, and $d_i S$ is the internal entropy produced inside the system when an irreversible process occurs. When this latter formulation of the second law is compared to the formulation of Nieuwenhuizen (equation (10)), it is possible to derive two relations, the first being:

$$d_e S = dS_{ep} + \frac{T_e}{T} dI \tag{16}$$

It is worth noticing that at equilibrium, $T_e = T$ and the external entropy contribution equals the total variation of entropy. The second relation concerns the entropy produced during an irreversible process:

$$d_i S = \frac{T - T_e}{T} dI \tag{17}$$

At equilibrium, the internal entropy contribution is equal to zero. When this equation above is compared with that obtained for the Fourier's law (equation (14)), the time derivative of the residual entropy of Nieuwenhuizen takes the simple following form:

$$\frac{dI}{dt} = \frac{K_i (T - T_e)}{T_e} \tag{18}$$

Now, in order to deeply establish the connection between this Fourier's approach and the Nieuwenhuizen's thermodynamics, let us return to the thermal model of the figure 1. From this picture, a quantity of heat supplied to the "phonon-bath" (fast modes) sub-system is separated into two parts following the energy balance equation:

$$\delta Q = C_\infty dT + C_i dT_e \tag{19}$$

This usual calorimetric equation reveals the heat capacities of the different parts of the entire system. The first term of the right-hand-side of this equation is related to the heat capacity of the infinitely fast degrees of freedom of the system all having the equilibrium temperature $T$. A point of some interest is that the quantity of heat is supplied via the surrounding only to them, which means that a heater (for example a Joule effect heater) is coupled principally to these rapid modes. The other part is related to the heat capacity of the slow internal degree of freedom, $C_i$, involved in the slow configurational change. A direct comparison with equation (9) yields to:

$$\begin{cases} dS_{ep} = C_{\infty} d(\ln T) \\ dI = C_i d(\ln T_e) \end{cases} \quad (20)$$

The preceding equations emphasize two features. Firstly, the equilibrium part of the entropy of Nieuwenhuizen is related to the heat capacity of the fast mode at equilibrium multiplied by the logarithmic variation of the classical temperature, such as described in classical thermodynamics. Secondly, it emphasizes the link between the configurational part of the entropy of Nieuwenhuizen with the heat capacity of the slow modes multiplied by the logarithmic variation of the effective temperature. Finally, it is trivial from (19) to obtain the following expression for the heat capacity:

$$C_P = C_{\infty} + C_i \frac{dT_e}{dT} \quad (21)$$

This expression of the heat capacity obtained with our thermal model included in the effective temperature framework of Nieuwenhuizen can now be compared with that provided in the section II.

**C. Relationships connecting $T_e$ and $\xi$**

In equalling the two expressions of the internal production of entropy (equations (2) and (17)) coming from the classical non-equilibrium thermodynamics and the effective temperature approaches, a first relationship connecting $\xi$ and $T_e$ is provided:

$$A\frac{d\xi}{dt} = (T - T_e)\frac{dI}{dt} \tag{22}$$

From this equation, we can say that the thermodynamic force (affinity) of the irreversible process driving the flux of the order parameter is equivalent to the departure of the fictive temperature from the equilibrium temperature driving the flux of the residual entropy of Nieuwenhuizen. By means of this previous relationship concerning the entropy production, and knowing that by definition $A = -\partial G/\partial \xi)_T = T\partial S/\partial \xi)_T - \partial H/\partial \xi)_T$, it is trivial to show that the equations (3) and (11) of the heat capacity expressed in the two different formalisms are equivalent, and particularly that:

$$\begin{cases} C_\xi = T\dfrac{dS_{ep}}{dT} \\ \partial H/\partial \xi)_T \dfrac{d\xi}{dT} = T_e \dfrac{dI}{dT} \end{cases} \tag{23}$$

As already mentioned, since the heat capacity is by essence able to stress how internal degrees of freedom can be thermally activated, the comparison can yield to interesting relations connecting the relaxing parameters coming from the different models. Particularly, in equalizing the expressions (5) and (21) of the heat capacities, a second relationship connecting the two parameters $\xi$ and $T_e$ is provided:

$$\partial H/\partial \xi)_T \frac{d\xi}{dt} = C_i \frac{dT_e}{dt} \qquad (24)$$

This equality occurs between two heat fluxes, one expressed in terms of the rate of the order parameter multiplied by the isothermal heat of the transformation (which are usual thermodynamic variables), and the other in terms of the heat capacity of the slow modes multiplied by the effective temperature rate (which have an existence only in the effective temperature approach). Let us mention that it has been implicitly assumed that $C_\infty = C_\xi$, which is trivial because these two quantities are related only to the infinitely fast degrees of freedom (glassy state). Finally, summing member per member this latter equation with (22), an interesting equation connecting the entropy flux due to the advance of the order parameter with the heat capacity of the slow modes multiplied by the effective temperature rate is obtained:

$$\partial S/\partial \xi)_T \frac{d\xi}{dt} = \sigma_i + \frac{C_i}{T}\frac{dT_e}{dt} \qquad (25)$$

In integrating this general entropy balance equation over a time interval at constant temperature, the entropy variation of the system due to the change of order can be obtained by means of the heat capacity of the slow modes taking into account the positive entropy produced over this time scale. Actually, this reveals that the variation of the entropy of the whole system can be regarded has an irreversible internal exchange of heat between two sub-systems of different temperature, not separated by spatial dimension but only by time.

**D. Relationships connecting $T_e$ and $\xi$ in the linear regime**

Now, let us consider the linear regime of the thermodynamics of irreversible processes. The fundamental assumption underlying the linear regime is to consider that the temperature rate is not too high in order that the moving variables are not too departed from their equilibrium values. It is the regime of validity of the so-called linear response theory. We are aware that in the case of the glass transition, when the kinetic relaxation times of the quantities involved in the process are very high, this assumption can bring some questions. However, let us envisage one consequence of our approach in this regime. As we have assumed that $\xi$ and $T_e$ are two variables coming from different frameworks describing the same underlying process, then near equilibrium their kinetic relaxation times must be equal. For example, near equilibrium, in considering that $\tau$ is the kinetic relaxation time constant, the order parameter relaxes toward its equilibrium value $\xi_{eq}$ following the simple relaxation equation:

$$\frac{d\xi}{dt} = -\frac{(\xi - \xi_{eq})}{\tau} \tag{26}$$

Consequently, from (24) and using the Tool's equation (8) with $\tau = 1/K_T$, we obtain:

$$\partial H / \partial \xi)_T^{eq} (\xi - \xi_{eq}) = C_i^{eq}(T_e - T) \tag{27}$$

Again, this latter relationship proves that any irreversible process can be regarded under two different, but however equivalent, points of view. The left-hand-side of (27) comes to consider heat exchange inside the system due to the departure from equilibrium of an order parameter. The right-hand-side comes to consider that this heat exchange holds within the system between the classical equilibrium temperature of the system and the effective temperature. Let us now derive a useful relation in this regime near equilibrium. In substituting $C_i^{eq}$ by its expression (second term of the right-hand member of (7) because $C_P^{eq} = C_\xi + C_i^{eq}$) and using the fact that, near equilibrium, $\partial H/\partial \xi)_T^{eq} \approx \partial H/\partial \xi)_T$ and also that the affinity can be developed to the first order in Taylor expansion of the extent of the transformation $A = -\partial G/\partial \xi)_T = -\partial^2 G/\partial \xi^2)_T^{eq}(\xi - \xi_{eq})$, then we obtain:

$$\frac{T_e}{T} = 2 - \frac{T\partial S/\partial \xi)_T}{\partial H/\partial \xi)_T} \qquad (28)$$

At $T = T_g$ (the glass transition temperature), this latter equation becomes:

$$\frac{T_e}{T_g} = 2 - \frac{1}{\Pi(T_g)} \qquad (29)$$

where $\Pi(T_g) = \dfrac{\partial H/\partial \xi)_{T=T_g}}{T_g \partial S/\partial \xi)_{T=T_g}} = \dfrac{\Delta C_P \Delta \kappa}{VT_g (\Delta \alpha)^2}$ is the so-called Prigogine-Defay ratio measured at $T_g$.[29] $\Delta C_P$ is the heat capacity jump, $\Delta \kappa$ the compressibility jump, $\Delta \alpha$ the thermal expansion coefficient jump near the glass transition temperature, and $V$ is the volume of the system. From (28), since $T_e$ is a time-dependent temperature, then the Prigogine-Defay ratio is time-dependent. For glassy system, however, we can assume that at the glass transition temperature the system becomes completely frozen and thus that $T_e$ becomes also progressively frozen. We will see in the next section that this frozen effective temperature which is actually the usual fictive temperature $T_f$ calculated from calorimetric experiments for example, is not a thermodynamic temperature. Consequently, the non-unity of the Prigogine-Defay ratio measured in experiments is due to the departure from $T_g$ of the fictive temperature and can be, to a first order, found from the simple equality (29) above. Another consequence is since $T_e$ can be greater or smaller than $T_g$, depending on the sign of the temperature rate, then the Prigogine-Defay ratio can be slightly greater or smaller than the unity (see for recent works on the Prigogine-Defay ratio the references [12,28,29]). In conclusion, we can see that this simple relation (29) holding only near equilibrium can be used to a first order to estimate from the measurement of $T_g$ and $T_f$ an approximate value of the the Prigogine-Defay ratio.

## IV. GENERALISATION OF THE NOTION OF TEMPERATURE FOR SYSTEMS OUT OF THERMODYNAMIC EQUILIBRIUM

### A. General definition of the effective temperature

In the previous section, we have supposed that $\xi$ and $T_e$ are not independent variables, but that they describe the same underlying phenomenon. We have provided different relationships connecting these variables from the expressions of the entropy and the heat capacities when these properties are expressed in the formalism of Tool/Nieuwenhuizen and in the formalism of De Donder/Prigogine/Defay. Certain of these relationships hold only in the linear regime. We would like now to push further the reasoning in considering simply that an effective temperature can be associated to any irreversible phenomenon existing in Nature. Indeed, since any irreversible process can be characterized by the advance of one or more order parameters in the course of time, generating thus a positive amount of entropy, then from the connection established previously, there is an effective temperature (or more) that characterizes thermodynamically the system undergoing this irreversible transformation. In other words, the time scale separation occurring between different internal degrees of freedom of a whole system gives rise to different thermodynamic temperatures inside this system. Precisely, from the equation (14) of the production of entropy in the effective temperature framework based on the Fourier's model, the following expression of the effective temperature is obtained:

$$\frac{T_e}{T} = 1 + \frac{\sigma_i}{2K_i}\left[1 \pm \left(1 + \frac{4K_i}{\sigma_i}\right)^{1/2}\right] \qquad (30)$$

The sign plus or minus means that the effective temperature can be greater or smaller than the real temperature of the system depending on the experimental conditions. Hence, it is sufficient that an evolving order parameter inside a system generates an amount of entropy to define an effective temperature different than the classical one. As a matter of generality, any irreversible process (whatever its nature) generating entropy gives rise to an effective temperature. Practically, in referring another time to the thermal picture of the figure 1, while this mentioned non-equilibrium event occurs, there is an exchange of heat between the effective temperature of the slow modes and the temperature of the fast modes within the system. Once the equilibrium is reached, the effective temperature does not exist anymore and the classical temperature of the system makes its classical sense. In fact, under this approach, the notion of thermodynamic equilibrium takes the same significance than the notion of thermal equilibrium.

Let us envisage simple practical examples to endorse our demonstration. Let be a diffusion of matter whenever a gradient of concentration of diffusing species holds inside the system. This diffusion of species (molecules, particles, ions, etc) is ruled by the Fick's law. It is an irreversible process, which generates entropy. Consequently, when a diffusion of matter happens inside a system due to gradient of concentration, then from (30) there is more than one macroscopic temperature characterizing the system. Let be a chemical reaction occurring in a system maintained at a classical constant temperature $T_0$. The production of entropy is given by $\sigma_i = A/T \times d\xi/dt$ where in this case $A$ is the chemical affinity of the process and $\xi$ the extent of the reaction. An intriguing consequence of our approach is that while this chemical reaction takes place at a well-defined classical temperature, the exact temperature of the system is actually not well-defined, and thus we must consider more than one thermodynamic temperature for the system. Finally, this general reasoning can be applied to all other irreversible phenomena (Ohm's law, thermoelectric effects, etc…). For example, when an amount of work is supplied irreversibly to a system, producing thus entropy, an effective temperature can be defined. In accordance with this approach, it may also be interesting to calculate the effective temperature in the case of coupled cross irreversible phenomena following the so-called Onsager's relations.

## B. The case of the glass transition

The glass transition is a very specific case of irreversible process for which the advance of the order parameter characterizing a structural change in the system becomes frozen at a temperature below $T_g$. We can summarize briefly under our approach what is expected for glasses:

-Well above $T_g$, although the system is under-cooled (in the case of a glass-forming liquid), there is no production of entropy because the relaxation time of the structural degree of freedom is too small; thus the system is considered at equilibrium and its temperature is well-defined.

-Around $T_g$, in the annealing range, there is production of entropy because of the relaxation of the structural degree of freedom (structural recovery); thus in this case, there is an effective temperature following (30) and the temperature of the system is not well defined.

-Lastly, below $T_g$, the system is completely glassy, the order parameter is completely frozen, there is no production of entropy and the temperature of the glass is well-determined. The glass is considered at thermodynamic equilibrium when considering time scales not too long. Under these circumstances, the temperature is uniquely determined by the energy fluctuations of the fast internal degrees of freedom. When the system is glassy, it is nevertheless possible to define a frozen effective temperature, which is the usual fictive temperature determined from calorimetric measurements for example.[35] It is the equilibrium temperature for the system when it starts to be glassy. But, this fictive temperature does not have the meaning of a real thermodynamic effective temperature which, under our approach, is uniquely defined when the system evolves and produces entropy. Another manner in envisaging these three cases, is to consider the gradual apparition of a time-scale separation between internal degrees of freedom whereas the temperature of the system is decreased following a defined cooling ramp. At high temperatures the time-scales of the several degrees of freedom are the same. At low temperatures for glasses, there are two well separated time-scales, one tending toward the infinity (frozen-in of the slow modes) and the other close to zero determining the thermodynamic temperature of the system. In the intermediary temperatures ranges, the two time-scales are not completely separated and the system has two thermodynamic temperatures. We will see now, how in changing the observation time scale, experimentalists can effectively access to the effective temperature.

# V. TEMPERATURE MODULATED CALORIMETRIC EXPERIMENTS REVEAL THE EFFECTIVE TEMPERATURE OF SYSTEMS OUT OF EQUILIBRIUM

As already mentioned, there are some questions concerning the possibility to have experimentally access to the effective temperature. One more time, we will show that heat capacity measurements can reveal that, in a system undergoing an irreversible physico-chemical transformation, there is an effective temperature at the same time that the classical temperature. This latter is the temperature which is measured with the thermometer because generally the thermometer and the heater are mostly coupled to the fast modes of the system. For instance, let us consider temperature modulated calorimetric experiments such as the famous Birge and Nagel specific heat spectroscopy experiments on glass-forming liquids.[44-46] Other type of temperature modulated experiments such as temperature modulated differential scanning calorimetry (TMDSC) or ac-calorimetry experiments can be envisaged.[47-50] In these types of calorimetric measurements, an oscillating thermal power is supplied onto a sample producing a harmonic temperature oscillation, $T_{ac} = \delta T_{ac} \exp(i\omega t - \varphi)$. The monitored temperature oscillation is directly linked to a frequency dependent complex heat capacity of the sample, $C^* = C' - iC''$. The great interest of these experimental methods is the existence of a well-defined observation time scale directly given by the period of the temperature modulation (or the frequency of the power modulation). This time-scale control parameter is a spectroscopic probe allowing the observance of the internal dynamic inside a sample undergoing a physico-chemical transformation. The principal interest of these modulated calorimetric methods as compared to classical DSC for glasses study comes from the stationary character of these methods (TMDSC not included). This means that the dynamic of the glass transition can be probed in changing the observation time scale at constant $\tau$ around a given temperature, while in scanning temperature methods it is the variation of $\tau$ with temperature which provokes the freezing-in of the system for a constant observation time scale.

The formalism used for deriving the complex heat capacity is usually the linear response theory. Nevertheless, Baur and Wunderlich have derived this dynamic property, in the linear regime, directly from macroscopic non-equilibrium thermodynamics.[51,52] From this latter approach, it is demonstrated that the imaginary part of the frequency dependent complex heat capacity is directly connected to the mean entropy production averaged over one period of the temperature cycle:

$$\Delta_i S = \oint \sigma_i dt = \pi \left( \frac{\delta T_{ac}}{T_0} \right)^2 C'' \qquad (31)$$

where $T_0$ is the mean dc temperature around which the temperature oscillation of amplitude $\delta T_{ac}$ occurs. Under the effective temperature formalism, from (14) we have:

$$\Delta_i S = \frac{K_i}{T_0} \oint \frac{(T - T_e)^2}{T_e} dt \qquad (32)$$

Hence, the measurement of an imaginary component in the frequency dependent complex heat capacity reveals that an oscillating effective temperature with amplitude attenuation and phase lag with respect to the driving oscillating temperature holds inside the sample. Since in modulated calorimetric experiments, there is only a mean production of entropy in a determined range around a given frequency $\omega_0 \approx 1/\tau(T_0)$, then in this range the temperature of the sample is not well-defined (see the graph of the figure 2). As depicted in the figure 2 where the imaginary part of the complex heat capacity is represented versus the thermal frequency, then at low frequencies the system remains at each instant in equilibrium. There is neither amplitude attenuation, nor phase lag of the effective temperature, which means that there is only one single temperature. At high frequencies, the integral in (32) tends toward zero, the effective temperature has no time to move over a period of the oscillation. The net entropy produced over this time scale is equal to zero. The system is glassy over this observation time scale, and its temperature which is only given by the energy repartition between the rapid internal degrees of freedom is well-defined. The expression of the frequency dependent complex heat capacity, has been either obtained by means of an oscillating effective temperature or by means of an oscillating order parameter, which confirms another time the equivalence of the two approaches.[53,54] Lastly, we have also recently shown that in modulated temperature calorimetric experiments when there is a relaxation time constant connected to the temperature of the sample (adiabaticity and thermal diffusivity) there is always the possibility to define an experimental frequency dependent complex heat capacity for which the imaginary part is connected to the averaged entropy produced over a period of the cycle.[55] Since, in the case of the usual frequency dependent complex heat capacity, the same relation exactly holds when an order parameter connected to a structural degree of freedom is involved, then it appears logical to define also a temperature (effective temperature) connected to the relaxation of this order parameter.

## VI. CONCLUSION

The new thermodynamics of Nieuwenhuizen based on an effective temperature as an extra thermodynamic variable has been explained under the framework of classical thermodynamics of irreversible processes. In comparing, on one hand, the equations of the entropy production and, in other hand, the equations of the heat capacity that can be obtained from the two approaches, several relationships have been derived connecting the effective temperature with the ordering parameter characterizing the departure from equilibrium of a structural internal degree of freedom inside a system. The connection between the two approaches has been strongly established in proposing a simple thermal picture defining the exchange of heat between two sub-parts inside a system possessing different temperatures, and where the equation of Fourier have been used.

From this point, the Fourier's model of the effective temperature, and particularly the entropy production calculated from this model, has been used in order to associate at each irreversible process existing in Nature an effective temperature. This is the generalisation of the macroscopic notion of temperature to systems out of thermodynamic equilibrium. From this point of view, the relaxation toward equilibrium of any internal degree of freedom inside a system when it is departed from equilibrium can be seen as relaxation of the effective temperature toward the classical bath temperature. In fact, what can be easily seen with this approach is that the classical notion of temperature is defined only when the energy exchanges within the system are equally distributed amongst the internal degrees of freedom constituting the system. When after a perturbation, some of the internal degrees of freedom are so slow that energy takes time to be transferred between them and from the fast ones the temperature of the whole system is not well-defined. There is not a single temperature inside the system. In this case, the slow modes have their proper temperature (effective temperature. Under this approach with the definition (30) of the effective temperature, the notion of thermodynamic equilibrium takes exactly the same meaning than thermal equilibrium.

What happens in the case of glasses? When the perturbation of the system is so fast that the structural degree of freedom has had not enough time to be excited, or in other words, that the order parameter has had not enough time to take a new value, then the system is frozen in a specific structural configuration (one macroscopic state) for which the representative entropy is thus equal to zero. Nevertheless in this case a frozen fictive temperature can be defined which can be estimated from a $H(T)$ or a $C_P(T)$ curve as it is usually done in glasses experiments. This is the equilibrium temperature that the system had at an anterior instant when it just began to be frozen. A certain quantity of energy is thus trapped in the frozen internal degree of freedom. But, since in this case the fast internal degrees of freedom continue to be equally excited, the temperature of the glass continue however to be well defined, but only by the energy fluctuations of these rapid modes. The frozen fictive temperature usually obtained in glass experiments does not have the meaning of a temperature. There is no production of entropy. It exists, however, intermediate situations for which the effective temperature is relaxing (aging effects), and where entropy is being produced, and where, as shown in this paper, there are several thermodynamic temperatures in the system.

The great interest of the heat capacity coefficient is its availability in revealing directly the quantity of heat trapped in frozen degrees of freedom or which is relaxing by thermalization process between the effective temperature and the bath temperature inside the system. The heat capacity is, actually, the ratio of the amount of heat supplied (or released) to the system from the exterior with the effectively recorded temperature variation (temperature of the rapid modes, with which the thermometer is coupled). Since, the quantity of heat supplied to the system is a priori well-defined, and that the infinitely fast degrees of freedom are all excited, the temperature elevation is equally the same, whether the slow modes are excited or not (temperature is intensive variable). In this case, the heat capacity is lower. In other words, the quantity of heat does not go where we believe that it should go, because a certain part is trapped somewhere inside the system. Indeed, in the case of glasses, the configurational heat capacity is equal to zero, and the measured contribution is always lower than the equilibrium one. In the intermediary regime, the trapped energy contribution becomes to be resumed, producing thus entropy (and disorder) whose signature is directly observable, for example, by the presence of an imaginary part in the frequency dependant complex heat capacity. During this relaxation process, there is a thermodynamic temperature other than that defined by the rapid modes inside the whole system, but due only to slow exchange of energy between the slow and fast modes.

This work was realized inside the "Thermodynamique des petits systèmes" team and the "Capteurs Thermométriques et Calorimétrie" pool of the "Institut Néel". The authors want to thank O. Bourgeois for numerous stimulating discussions and V. Forge for judicious comments and several corrections in reading the manuscript.

Figure 1: Thermal model representing a thermodynamic system composed by two distinctive parts, one being composed by the fast degrees of freedom of the system (phonon bath) with the heat capacity $C_\infty$ and the usual equilibrium temperature $T$, and the other composed by the slow internal degrees of freedom with the heat capacity $C_i$ and the effective temperature $T_e$. The "$C_i$" heat capacity part is thermally coupled by a thermal conductance $K_i$ to the "$C_\infty$" one, this latter being thermally coupled to the outside world (thermal bath) by a thermal conductance $K_0$. The quantity of heat $\delta Q$ is only supplied to the fast modes of the system.

Figure 2: Mean entropy calculated over one cycle of the temperature oscillation during a temperature modulated calorimetric experiment versus the frequency of the temperature oscillation. We can find two ranges where the mean entropy produced is equal to zero. One for frequencies tending toward the infinity (glassy state) and the other for frequencies tending towards zero (true equilibrium state). For these two cases, since the mean entropy produced is equal to zero, the system is considered at equilibrium and the temperature is well-defined, although for the glassy state some of the internal degrees of freedom have been frozen-in. In the frequency range where the mean entropy produced is different from zero, the temperature is not well-defined and there is the presence of a frequency dependent effective temperature.

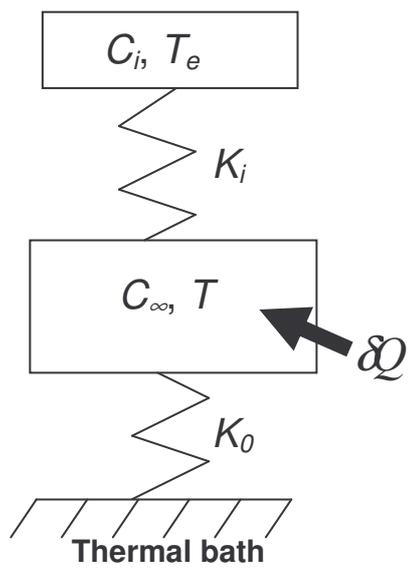

J.-L. Garden : Figure 1
Journal of Chemical Physics

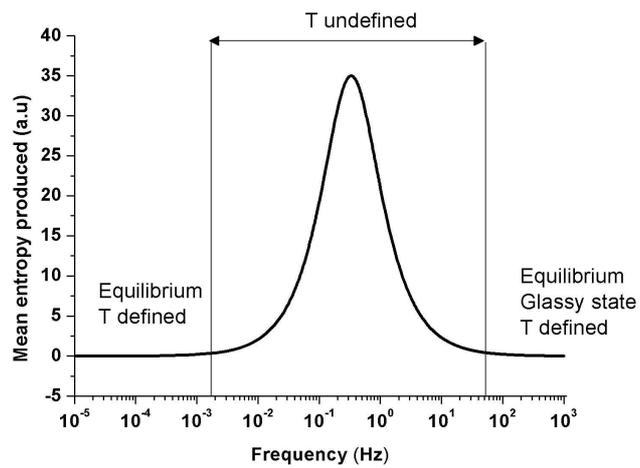

J.-L. Garden Fig 2
Journal of Chemical Physics